\documentclass{llncs}
\pdfoutput=1
\usepackage[T1]{fontenc}

\usepackage{makeidx}

\usepackage{makeidx}
\usepackage{graphicx}
\usepackage{subfigure}
\usepackage{amssymb}
\usepackage{amsmath}
\usepackage{amscd}

\newcommand{\commentout}[1]{}

\newcommand{\R}{\mathbb{R}}                    

\newcommand{\abs}[1]{\mathop{\left\lvert #1 \right\rvert}} 
\newcommand{\args}[1]{\mathop{\left( #1 \right)}}

\newcommand{\cbrace}[1]{\mathop{\left\{ #1 \right\}}}
\newcommand{\bracket}[1]{\mathop{\left[ #1 \right]}}

\DeclareMathOperator{\wdeg}{wdeg}       

\renewcommand{\S}[1]{{\mathcal{#1}}}           

\newcommand{\widebar}[1]{\overline{\!#1}}   

\newcommand{\atab}[1]{\hspace*{#1em}}

\newcounter{algorithm_counter}
\newenvironment{algorithm}[1]{
\refstepcounter{algorithm_counter}
\setlength{\parindent}{0\parindent}
\vspace{2ex}
\begin{minipage}{\textwidth}
\rule{\textwidth}{5\arrayrulewidth}\\
\begin{footnotesize}
{\bf Algorithm \arabic{algorithm_counter}} (#1) \\
\rule[+1.5ex]{\textwidth}{\arrayrulewidth}

\vspace{-1.5ex}

}%
{
\\[-1.5ex]
\rule{\textwidth}{\arrayrulewidth}
\end{footnotesize}
\end{minipage}
\setlength{\parindent}{\parindent}
}

{
\\[-1.5ex]
\rule{\textwidth}{\arrayrulewidth}
\end{footnotesize}
\end{minipage}
\setlength{\parindent}{\parindent}
}

\begin{document}
\title{Extending Bron Kerbosch for the Maximum Weight Clique Problem}
\titlerunning{Bron Kerbosch}

\author{Brijnesh J.~Jain$^1$  \and Klaus Obermayer$^1$} 
\authorrunning{Brijnesh J.~Jain and Klaus Obermayer}  
\institute{$^1$Berlin Institute of Technology, Germany\\
\email{\{jbj|oby\}@cs.tu-berlin.de}}

\maketitle 

\begin{abstract}  
This contribution extends the Bron Kerbosch algorithm for solving the maximum weight clique problem, where continuous-valued weights are assigned to both, vertices and edges. We applied the proposed algorithm to graph matching problems.
\end{abstract}

\section{Introduction}

Comparing structural variations of two graphs is a fundamental task in pattern recognition, which finds its applications in diverse areas such as computer vision, bioinformatics, and computational chemistry.

Since the graph matching problem is well-known to be NP-hard, there is an ongoing research on devising optimal and approximate graph matching algorithms.
One popular technique consists in transforming graph matching to an equivalent clique search in a derived auxiliary structure, called association graph \cite{Ambler73,Barrow76,Bartoli00,Pelillo99a,Pelillo99b,Pelillo99c,Raymond02}. In \cite{Jain09}, it has been shown that a broad range of graph matching problems can be reduced to the maximum weight clique problem. Examples include divers graph distances and matching problems such as the graph edit distance, dissimilarities based on the maximum common subgraph, hierarchical tree matching, geometric graph distance functions, and many-to-many graph matching problems, to mention a few. 

A maximum weight clique is a clique of a weighted graph with maximum sum of vertex and edge weights. The problem is that clique search algorithms are only well investigated for graphs without weights or graphs with vertex weights, only. 

In this paper, we present an extension of the Bron-Kerbosch algorithm \cite{Bron73} for solving the maximum weight clique problem where the underlying graph has continuous-valued weights assigned to both, vertices and edges. We present und discuss first experiments.

\section{Preliminaries}

\subsection{Attributed Graphs}
Let $\S{A}$ be a set of \emph{attributes} and let $\varepsilon \in \S{A}$ be a distinguished element denoting the \emph{null} or \emph{void} element.  An  \emph{attributed graph} is a tuple $X = (V, \alpha)$ consisting of a finite nonempty set $V$ of \emph{vertices} and an \emph{attribute function} $\alpha: V \times V \rightarrow \S{A}$. Elements of the set $E = \cbrace{(i,j)\in V \times V \,:\, i \neq j \text{ and }\alpha(i,j) \neq \varepsilon}$ are the \emph{edges} of $X$. 

In this definition, attributes assigned to vertices $i \in V$ are given by $\alpha(i,i)$ and edges are characterized by pairs of distinct vertices that have non-null attributes. The vertex set of an attributed graph $X$ is often referred to as  $V_X$ and its attribute function as $\alpha_X$.  

A \emph{subgraph} of $X$ is a graph $Y$ with vertex set $V_Y \subseteq V_X$ and attribute function $\alpha_Y(i,j) \in \cbrace{\alpha_X(i,j), \varepsilon}$ for all $i,j \in V_Y$. We write $Y \subseteq X$ to denote that $Y$ is a subgraph of $X$. An \emph{induced subgraph} of $X$ is a subgraph $Y \subseteq X$ with attribute function $\alpha_Y = \alpha_{|V_Y}$. We write $X[V_Y]$ to denote the subgraph of $X$ induced by the vertex set $V_Y$. 

A graph is said to be \emph{complete} if all of its vertices are mutually connected by an edge. A \emph{clique} of a graph $X$ is a subset $C \subseteq V_X$ such that the induced subgraph $X[C]$ is complete. A clique $C$ of $X$ is said to be \emph{maximal} if $C$ is not contained in any larger clique of $X$. A \emph{maximum clique} is a clique of $X$ with maximum cardinality of vertices. 

The set $N(i) = \cbrace{j \in V_X \,: \, (i,j) \in E}$ defines the set of all vertices of $X$ adjacent to $i \in V_X$. Note that the set $N(i)$ excludes vertex $i$. The number $\deg(i) = \abs{N(i)}$ is the \emph{degree} of vertex $i\in V_X$.  

Suppose that $Y$ is a subgraph of $X$. The deletion of $Y$ in $X$ is defined by the graph $Z = X-Y$ with vertex set $V_Z = V_X$ and attribute function 
\[
\alpha_Z(i,j) = \begin{cases}
\varepsilon &: \quad i,j \in V_Y\\
\alpha_X(i,j) &: \quad\mbox{otherwise}
\end{cases}
\]

Let $X$ and $Y$ be graphs. A \emph{partial morphism} from $X$ to $Y$ is a partial injective mapping
\[
\phi : V_X \rightarrow V_Y, \quad i \mapsto i^\phi.
\]
By $\S{D}(\phi) \subset V_X$ we denote the \emph{domain} of $\phi$ . A \emph{morphism} is a partial morphism $\phi$ which can not be extended to a partial morphism $\phi'$ with larger domain, that is $\S{D}(\phi) \subsetneq \S{D}(\phi')$. By $\S{M}(X,Y)$ we denote the set of all morphisms from $X$ to $Y$.
\subsection{The Maximum Weight Clique Problem}

A \emph{weighted graph} is a graph $Z = (V, \alpha)$, where the underlying attribute set is of the form $\S{A} = \R \cup \cbrace{\varepsilon}$. An \emph{unweighted graph} is a weighted graph $Z$ with attribute set of the form $\S{A} = \cbrace{0, 1}$ and  an attribute function $\alpha_Z$ that assigns each vertex the value 0 and each edge the value $1$ as its attribute. Thus, the distinguished null attribute $\varepsilon$ is represented by the value $0$. 

Suppose that $i \in V_Z$ is a vertex of a weighted graph $Z$. The \emph{weighted degree} of $i$ is defined by
\[
\wdeg_Z(i) = \alpha_Z(i,i) \; + \sum_{j \in N(i)} \alpha_Z(i,j).
\]
Note that in the case of unweighted graphs the notion of degree and weighted degree coincide.

The \emph{weight} of a clique $C$ of $Z$ is defined by
\[
\omega(C) = \sum_{i,j\in C} \alpha(i,j)
\]
The weight of a clique $C$ is the total sum of all vertex and edge weights of the induced subgraph $Z[C]$. Since the vertices of $Z[C]$ are mutually adjacent, the null attribute $\varepsilon$ does not occur in the definition of $\omega(C)$.

A \emph{maximum weight clique problem} is a combinatorial optimization problem of the form
\begin{align*}
\mbox{maximize} &\quad \omega(C)\\
\mbox{subject to} &\quad C \in \S{C}(Z),
\end{align*}
where $\S{C}(Z)$ is the set of all cliques of $Z$. Any solution of the maximum weight clique problem is a maximum weight clique of $Z$. A \emph{maximal weight clique} of $Z$ is a clique $C$ of $Z$ such that 
\[
C \subseteq C' \quad \Rightarrow \quad \omega(C) \geq \omega(C')
\]
for all cliques $C'$ of $Z$. It is impossible to enlarge a maximal weight clique $C$ to a clique $C'$ with higher weight. If all vertices and edges of $Z$ are associated with positive weights, a maximal weight clique is not a proper subset of another clique.

\subsection{Graph Matching as Clique Search}

To measure the structural variation of two graphs, we consider the following (indefinite) graph kernel
\[
k(X,Y) = \max_{\phi\in \S{M}(X,Y)} \sum_{i,j \in \S{D}(\phi)} k_{\S{A}}\args{\alpha_X(i,j), \alpha_Y(i^\phi, j^\phi)},
\]
where $k_\S{A}: \S{A} \times \S{A} \rightarrow \R_+$ is a positive definite kernel defined on the set $\S{A}$ of attributes. The graph kernel $k(X,Y)$ induces the notion of a well-defined \emph{length} of a graph $X$ by
\[
l(X) = \sqrt{k(X, X)}.
\]
As shown in \cite{Jain09b}, the graph kernel together with the length of a graph satisfy the Cauchy-Schwarz inequality
\[
\abs{k(X,Y)} \leq l(X)\cdot l(Y).
\]

Suppose that $X$ and $Y$ are two attributed graphs. We transform the problem of computing the graph kernel $k(X,Y)$ to the maximum weight clique problem of an association graph of $X$ and $Y$. An association graph $Z = X \otimes Y$ of $X$ and $Y$ consists of a vertex set $V_Z = V_X \times V_Y$ and an attribute function of the form
\[
\alpha_Z: V_Z \times V_Z \rightarrow R, \quad \big((i,j), (r,s)\big) \mapsto k_{\S{A}}\big((i,j), (r,s)\big).
\]
As shown in \cite{Jain09}, there is a one-to-one correspondence between the optimal solutions of the graph matching problem and the maximum weight cliques of $Z$.

\section{Extension of Bron-Kerbosch to the MWCP}  

\subsection{Basic Bron-Kerbosch}

\begin{figure}[tbp]
\begin{algorithm}{Basic Bron-Kerbosch for enumerating all maximal weight cliques}\label{alg:basic-bk}
01 \atab{0} call: $bk(\emptyset, V, \emptyset)$\\[0.5ex]
02 \atab{0} $bk(C, P, S)$ \\[0.5ex]
03 \atab{2} \textbf{if} $P = \emptyset$ and $S = \emptyset$ \textbf{then}\\[0.5ex]
04 \atab{4} report maximal weight clique $C$\\[0.5ex]
05 \atab{2} \textbf{for} each vertex $i \in P$ \textbf{do}\\[0.5ex]
06 \atab{4} $bk\Big(C \cup \cbrace{i}, P \cap N(i), S \cap N(i)\Big)$\\
07 \atab{4} $P = P \setminus \cbrace{i}$\\[0.5ex]
08 \atab{4} $S = S \cup \cbrace{i}$
\end{algorithm}
\end{figure}

Extension of the Bron-Kerbosch algorithm from enumerating all maximal cliques of an unweighted graph to enumerating all maximal weight cliques of a weighted graph is straightforward, since the notions of maximal clique and maximal weighted clique coincide for graphs with positive weights. Algorithm \ref{alg:basic-bk} outlines the standard Bron-Kerbosch procedure for enumerating the maximal weight cliques of a given graph $Z = (V, \alpha)$. The algorithm operates on three disjoint subsets $C$, $P$, and $S$ of  vertices from $V$. The set $C$ contains the vertices belonging to the current clique. Set $P$ maintains all prospective vertices, each of which is connected to all vertices of $C$. Vertices from $P$ are used for expanding the current clique $C$. Finally, the set $S$ contains all vertices that can no longer be used for completion of $C$, because all maximal cliques containing these vertices have already been reported. The Bron-Kerbosch algorithm is called with $C = S = \emptyset$ and $P = V$.

\subsection{Bron-Kerbosch with Pivoting}

In the case of unweighted graphs, the standard Bron-Kerbosch procedure described in Algorithm \ref{alg:basic-bk} is inefficient in the case of graphs with many non-maximal cliques. Bron and Kerbosch \cite{Bron73} introduced a variant of the standard algorithm involving a \emph{pivot vertex} $i_p$ chosen from $P$.\footnote{As shown by \cite{Koch00}, the pivot vertex can be more generally chosen from $P \cup S$. We do not consider this case here.} Any maximal clique of $S$ either includes the pivot vertex $i_p$ or one of the vertices $i \in P\setminus N(i_p)$ not adjacent to $i_p$. Therefore, only the pivot vertex $i_p$ and vertices from $P$ not adjacent to $i_p$ need to be considered as expansions of the current clique $R$ in each recursive call of the Bron-Kerbosch algorithm. Vertices $i$ from $P$ adjacent to $i_p$ can be skipped, because any clique containing $i$ must also contain $i_p$. Such a clique will be discovered in a subsequent recursive call once $i_p$ has been added to $C$. Algorithm \ref{alg:pivot-bk} presents the Bron-Kerbosch procedure with pivoting for enumerating all maximal weight cliques of $Z$.

\begin{figure}[tbhp]
\begin{algorithm}{Bron-Kerbosch with pivoting}\label{alg:pivot-bk}
01 \atab{0} call: $bk(\emptyset, V, \emptyset)$\\[0.5ex]
02 \atab{0} $bk(C, P, S)$ \\[0.5ex]
03 \atab{2} \textbf{if} $P = \emptyset$ and $S = \emptyset$ \textbf{then}\\[0.5ex]
04 \atab{4} report maximal weight clique $C$\\[0.5ex]
05 \atab{2} choose pivot vertex $i_p \in P$\\[0.5ex]
06 \atab{2} \textbf{for} each vertex $i \in P \setminus N(i_p)$ \textbf{do}\\[0.3ex]
07 \atab{4} $bk\Big(C \cup \cbrace{i}, P \cap N(i), S \cap N(i)\Big)$\\
08 \atab{4} $P = P \setminus \cbrace{i}$\\[0.5ex]
09 \atab{4} $S = S \cup \cbrace{i}$
\end{algorithm}
\end{figure}

The challenge of Bron-Kerbosch with pivoting consists in finding good pivot selection strategies. In the case of unweighted graphs different strategies have been suggested (see e.g. \cite{Koch00 and references therein}). For the more general case of weighted graphs, we suggest the following pivot selection strategies:
\begin{enumerate}
\item \emph{Random selection}: \\
The pivot vertex $i_p$ is randomly chosen from the set $P$.
\item \emph{Max-weighted degree selection}: \\
The pivot vertex $i_p$ is chosen from $P$ according to the rule
\[
\wdeg_{Z[C \cup P]}(i_p) \geq \wdeg_{Z[C \cup P]}(i)
\]
for all $i \in P$, where the weighted degree is taken with respect to the subgraph $Z[C \cup P]$ induced by the vertices of $C \cup P$.
\item \emph{Max-weight clique selection}: \\
The pivot vertex $i_p$ is chosen from $P$ according to the rule
\[
\omega\Big(C \cup \cbrace{i_p}\Big) \geq \omega\Big(C \cup \cbrace{i}\Big)
\]
for all $i \in P$.
\end{enumerate}

\subsection{Bron-Kerbosch for Solving the MWCP}

Often it is sufficient to report a single maximum weight clique rather than enumerating all maximal weight cliques. In this case, we modify Algorithm \ref{alg:pivot-bk} by recording the maximal weight clique $R_*$ with maximum weight found so far. To improve efficiency, we introduce a function $h:\S{C}(Z) \rightarrow \R_+$ with the following property:
\[
\forall C, C' \in \S{C}(Z): \quad C \subseteq C'  \;\Rightarrow \; \omega(C) + h(C) \geq \omega(C')
\]
Similarly, as in the $A^*$-algorithm, the function $h$ estimates the total weight obtained by expanding the current clique to a maximal clique. We demand that $\omega(C) + h(C)$ always overestimates the total weight of any clique containing $C$.

Algorithm \ref{alg:bk2} outlines the Bron-Kerbosch procedure for solving the MWCP. The set $C_*$ is a global variable which can be initialized by the empty set or an arbitrary clique of $Z$.

\begin{figure}[tbhp]
\begin{algorithm}{Bron-Kerbosch for Solving the MWCP}\label{alg:bk2}
01 \atab{0} initialize $C_*$\\[0.5ex]
02 \atab{0} call: $bk(\emptyset, V, \emptyset)$\\[0.5ex]
03 \atab{0} $bk(C, P, S)$ \\[0.5ex]
04 \atab{2} \textbf{if} $P = \emptyset$ and $S = \emptyset$ \textbf{then}\\[0.5ex]
05 \atab{4} \textbf{if}  $\omega(C) > \omega(C_*)$ \textbf{then}\\[0.5ex]
06 \atab{6} $C_* = C$\\[0.5ex]
07 \atab{2} \textbf{if} $\omega(C) + h(C) > \omega(C_*)$ \textbf{then}\\[0.5ex]
08 \atab{4} choose pivot vertex $i_p \in P$\\[0.5ex]
09 \atab{4} \textbf{for} each vertex $i \in P \setminus N(i_p)$ \textbf{do}\\[0.3ex]
10 \atab{6} $bk\Big(C \cup \cbrace{i}, P \cap N(i), S \cap N(i)\Big)$\\
11 \atab{6} $P = P \setminus \cbrace{i}$\\[0.5ex]
12 \atab{6} $S = S \cup \cbrace{i}$
\end{algorithm}
\end{figure}

Besides finding a good strategy for selecting the pivot vertex, a challenge for improving the efficiency of Algorithm \ref{alg:bk2} consists in formulating a good estimate function $h$. We suggest the following estimate functions $h$ for the maximum weight clique problem of $Z$.

\begin{enumerate}
\item \emph{Max-weight degree estimate (deg)}:\\ 
\[
h_{\deg}:\S{C}(Z) \rightarrow \R_+, \quad C \mapsto \max_{i \in P} \wdeg(i)
\]

\item \emph{Cauchy-Schwarz estimate (cs)}:\\
We assume that $Z = X \otimes Y$ is an association graph of $X$ and $Y$. Suppose that $X_C \subseteq X$ and $Y_C \subseteq Y$ are the induced subgraphs obtained by projecting the current clique $C$ to the factor graphs $X$ and $Y$. Let $\widebar{X}_C = X - X_C$ and $\widebar{Y}_C = Y - Y_C$ denote the graphs obtained by deleting $X_C$ in $X$ and $Y_C$ in $Y$. Then we have
\[
\abs{k\args{\,\widebar{X}_C, \widebar{Y}_C}} \leq  l\args{\,\widebar{X}_C} \cdot \, l\args{\,\widebar{Y}_C} 
\]
according to the Cauchy-Schwarz inequality. Thus, the estimate function
\[
h_{\mbox{cs}}:\S{C}(Z) \rightarrow \R_+, \quad C \mapsto l\args{\,\widebar{X}_C} \cdot \, l\args{\,\widebar{Y}_C} 
\]
overestimates the remaining weights of any maximal weight clique of $Z$ containing $C$.
\end{enumerate}
The deg-estimate is more general than the cs-estimate and can be applied to the generic maximum weight clique problem. In contrast, the cs-estimate is only applicable for graph matching problems that calculate geometric graph distance functions, that is graph distance functions that are maximizers of inner products. Since $h_{\deg} \leq h_{\mbox{cs}}$, our choice is the cs-estimate in case of its applicability.

\section{Experiments}

We applied the extended Bron Kerbosch algorithm to the problem of graph matching in order to assess its performance and to investigate its behavior. The aim is to investigate the effects of  different pivoting strategies and compared the matching performance of Bron Kerbosch with the graduated assignment algorithm \cite{Gold96}.

\subsubsection{Data.}

For this, we selected the following data sets from the IAM graph database repository: letter, grec, coil, and mutagenesis. We used the whole training sets of the letter and grec. For coil and mutagenesis, we considered the first $200$ graphs of the respective training sets. Table \ref{tab:characteristics} provides a summary of the main characteristics of the data sets. For further details we refer to \cite{Riesen08}. 

\begin{table}[t]
\centering
\begin{tabular}{l@{\qquad}ccccc}
\hline
\hline
data set & \#(classes) & avg(nodes) & max(nodes) & avg(edges) & max(edges) \\
\hline
letter &  15 & 4.7 & 8 & 3.1 & 6\\
grec &  22 & 11.5 & 24 & 11.9 & 29\\
coil &  100 & 8.3 & 26 & 14.1 & 48\\
molecules & 2 & 24.6 & 40 & 25.2 & 44\\
\hline
\hline
\end{tabular}
\vspace{1ex}
\caption{Summary of main characteristics of the data sets. The graphs were taken from the respective training sets.}
\label{tab:characteristics}
\end{table}

\newcommand{\figwidth}{0.48}
\begin{figure}%
\centering
\subfigure[letter - similarity]{\includegraphics[width=\figwidth\textwidth]{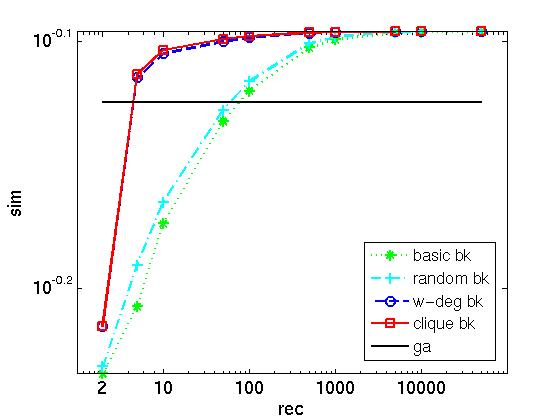}}\hfill
\subfigure[letter - time]{\includegraphics[width=\figwidth\textwidth]{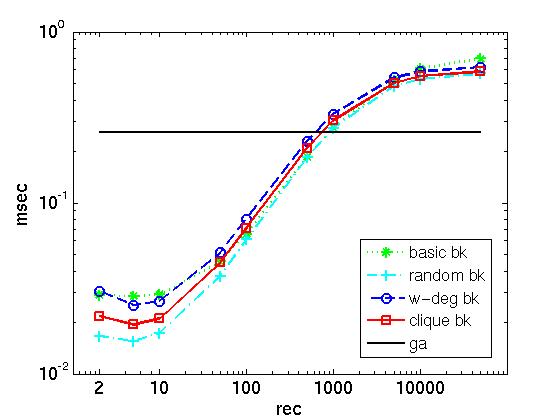}}\\
\subfigure[grec  - similarity]{\includegraphics[width=\figwidth\textwidth]{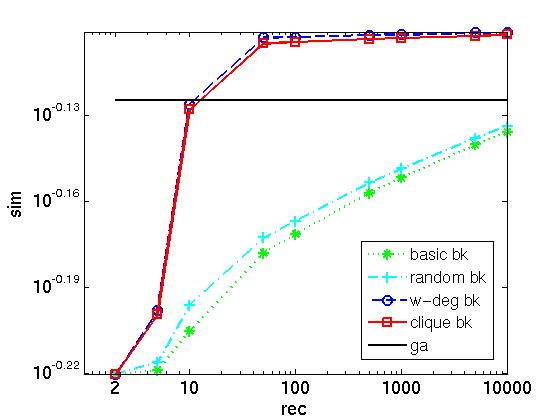}}\hfill
\subfigure[grec - time]{\includegraphics[width=\figwidth\textwidth]{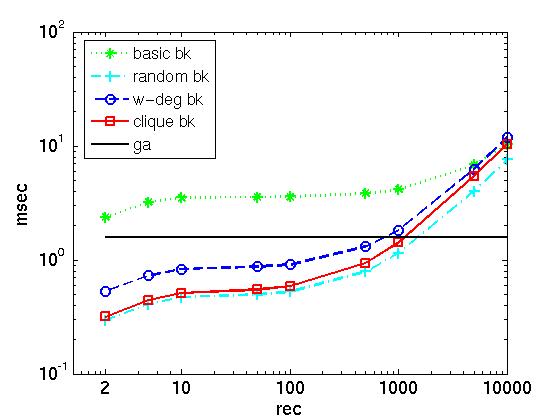}}\\
\subfigure[coil  - similarity]{\includegraphics[width=\figwidth\textwidth]{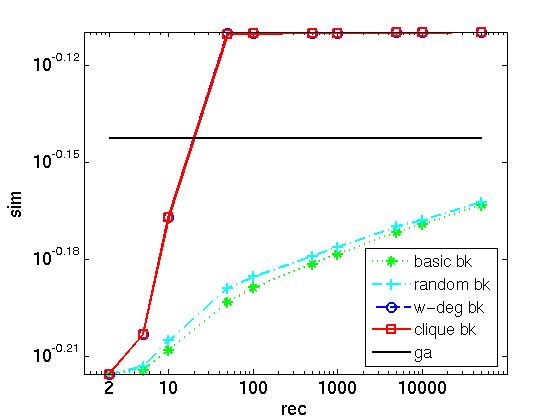}}\hfill
\subfigure[coil - time]{\includegraphics[width=\figwidth\textwidth]{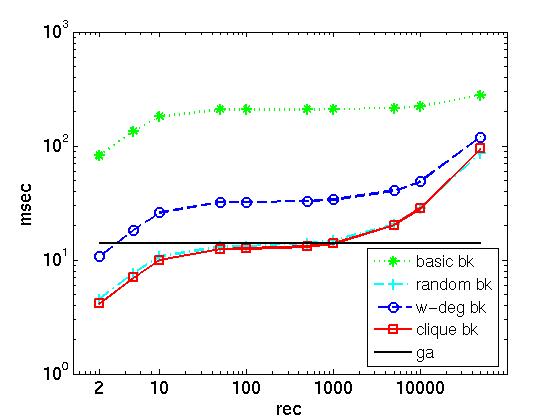}}\\
\subfigure[mutagenesis  - similarity]{\includegraphics[width=\figwidth\textwidth]{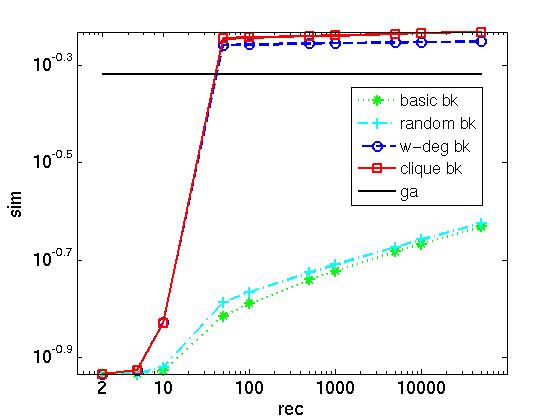}}\hfill
\subfigure[mutagenesis - time]{\includegraphics[width=\figwidth\textwidth]{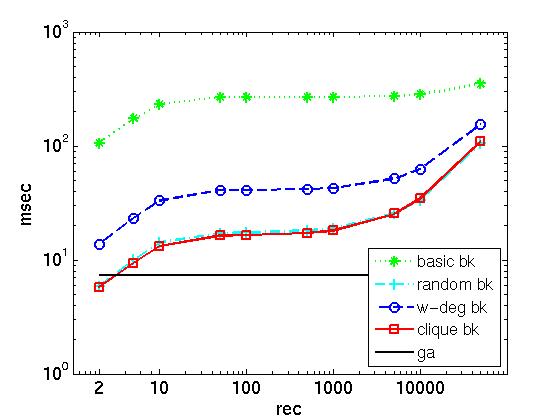}}
\caption{Results of Graduated Assignment (ga) and the $4$ Bron Kerbosch (bk). Shown are the average similarity and  computation time as a function of the number of recursions.}
\label{fig:results:matching}
\end{figure}

\subsubsection{Protocol.}
For each data set, we computed the pairwise similarities
\[
s(X,Y) = \frac{k(X, Y)}{l(X)l(Y)} \in \bracket{-1,+1}.
\]
For calculating pairwise similarities, we applied four variants of Bron Kerbosch. The variants differ in the choice of the following pivoting strategies: basic (no pivoting), random selection, w-deg selection, and clique selection. All four variants of Bron Kerbosch used the Cauchy-Schwarz estimate. We recorded the average similarity and computation time of the different variants of Bron Kerbosch after $\alpha$ recursive calls, where 
\[
\alpha \in \cbrace{2, 5, 10, 50, 100, 500, 1\,000, 5\,000, 10\,000, 50\,000}.
\]

\subsubsection{Results.}

Figure \ref{fig:results:matching} summarizes the results. From the plots we see that Bron Kerbosch with pivoting (random, w-deg, clique selection) is on average faster and scales better with problem size than Bron Kerbosch without pivoting (basic). This behavior is in line with findings of the standard Bron Kerbosch algorithm for the unweighted maximum clique problem. In addition, Bron Kerbosch using w-deg and clique selection outperform Bron Kerbosch using first and random selection with respect to solution quality. The solution quality of Bron Kerbosch with w-deg and clique selection are comparable. Bron Kerbosch with w-deg selection, however, is computationally more demanding than Bron Kerbosch with clique selection for two reasons: (i) w-deg selection needs significantly more recursive calls, and (ii) for each recursive call, w-deg selection is computationally more expensive than clique selection.  These findings make clique selection as our first choice for selecting the next pivot vertex.

Comparing the extended Bron Kerbosch algorithm using clique selection with graduated assignment shows that Bron Kerbosch returns significantly better results than graduated assignment in less time for letter, grec, and coil. For mutagenesis, graduated assignment provides a superior trade-off between speed and accuracy.

 \section{Conclusion}
 
 The extended Bron Kerbosch algorithm solve the maximum weight clique problem, where continuous-valued weights are assigned to both, vertices and edges. In doing so, the proposed algorithm is a generic tool for efficiently solving a broad range of graph matching problems. 
 
 Further research aims at applying Bron Kerbosch to classification and clustering problems in the domain of graphs. In addition, we are interested to which extent the Cauchy-Schwarz estimate improves Bron Kerbosch using different pivoting strategies.
 
 \bibliographystyle{splncs}

\end{document}